\documentclass[prl,aps,twocolumn,showpacs,floatfix]{revtex4}

\usepackage[dvips]{graphicx}
\usepackage{amsmath}
\usepackage{amssymb}
\usepackage{natbib}

\newcommand{\betam}{\beta_{\mathrm{max}}}
\newcommand{\Tm}{T_{\mathrm{max}}}
\newcommand{\Veff}{V_{\mathrm{eff}}}
\newcommand{\e}{e}%{\mathrm{e}}

\begin{document}
\title{Bimodality and hysteresis in systems driven by confined L\'evy flights}

\author{Bart{\l}omiej Dybiec}
\email{bartek@th.if.uj.edu.pl}
\affiliation{M. Smoluchowski Institute of Physics, and Mark Kac Center for Complex Systems Research, Jagellonian University, ul. Reymonta 4, 30--059 Krak\'ow, Poland}

\author{Ewa Gudowska-Nowak}
\email{gudowska@th.if.uj.edu.pl}
\affiliation{M. Smoluchowski Institute of Physics, and Mark Kac Center for Complex Systems Research, Jagellonian University, ul. Reymonta 4, 30--059 Krak\'ow, Poland}

\date{\today}

\begin{abstract}
We demonstrate occurrence of bimodality and dynamical hysteresis
in a system describing an overdamped quartic oscillator perturbed
by additive white and asymmetric L\'evy noise. Investigated
estimators of the stationary probability density profiles display
not only a turnover from unimodal to bimodal character but also a
change in a relative stability of stationary states that depends
on the asymmetry parameter of the underlying noise term. When
varying the asymmetry parameter cyclically, the system exhibits a
hysteresis in the occupation of a chosen stationary state.
\end{abstract}


\maketitle

%%%%%%%%%%%%%%%%%%%%%%%%%%%%%%%%%%%%%%%%%%%%%%%%%%%%%%%%%%%%%%%%%%%%%%%%%%%%%%%%%%%%%%%%%
%%
%% INTRODUCTION
%%
\section{Introduction}
The Langevin description of an overdamped Brownian motion in a potential $V(x)$
\begin{equation}
\dot{x}(t) = -V'(x)+\zeta(t),
\label{eq:lang}
\end{equation}
constitutes a basic paradigm to study the effects of fluctuations at the
mesoscopic scale \cite{vankampen1981,horsthemke1984,gammaitoni1998}. Here prime
stands for differentiation over $x$ and $\zeta(t)$ is a commonly assumed white
Gaussian noise representing close-to-equilibrium fluctuations of the dynamic
variable $x$. In contrast, in far-from-equilibrium situations, the Gaussianity
of the noise term may be questionable, due to e.g. strong interaction with the
surrounding ``bath'' \cite{klimontovich1994,schlesinger1995}. A natural
generalizations to the Brownian motion are then $\delta$-correlated L\'evy
stable processes which can be interpreted as fluctuations resulting from strong
collisions between the test particle and the environment. In particular, the
scale-free, self similar feature of L\'evy distributions gives rise to the
occurrences of large increments of the position coordinates $\Delta x$ during
small time intervals causing a non-local character of the motion. Within the
paper we address the problem of kinetics as described by Eq.~(\ref{eq:lang})
under the action of $\zeta(t)$ representing a stationary white L\'evy noise
\cite{janicki1994,dybiec2007}. Accordingly, the position of the Brownian
particle subjected to additive white L\'evy noise is calculated by direct
integration of Eq.~(\ref{eq:lang}) with respect to the $\alpha$-stable measure \cite{janicki1994,janicki1996,dybiec2004,dybiec2004b,dybiec2006,dybiec2007} $L_{\alpha,\beta}(s)$, i.e.,
$
x(t+\Delta t) = x(t)-V'(x(t))\Delta t + (\Delta t)^{1/\alpha}\zeta,
$
where $\zeta$ is distributed according to the $\alpha$-stable L\'evy type
distribution $L_{\alpha,\beta}(\zeta;\sigma,\mu=0)$ whose representation
\cite{feller1968,janicki1994,janicki1996} is given by the characteristic
function $\phi(k)$ defined in the Fourier space $\phi(k) =
\mathcal{F}(L_{\alpha,\beta}(\zeta;\sigma,\mu))= \int_{-\infty}^{\infty} d\zeta
\e^{ik\zeta} L_{\alpha,\beta}(\zeta;\sigma,\mu)$ for $\alpha\ne1$

\begin{equation}
\phi(k) = \exp\left[ -\sigma^\alpha|k|^\alpha\left( 1-i\beta\mathrm{sgn}(k)\tan
\frac{\pi\alpha}{2} \right) +i\mu k \right],
\end{equation}
and for $\alpha=1$
\begin{equation}
\phi(k) =
\exp\left[ -\sigma|k|\left( 1+i\beta\frac{2}{\pi}\mathrm{sgn} (k) \ln|k| \right) + i\mu k \right].
\label{eq:charakt}
\end{equation}
The stability index $\alpha$, determining tails of the probability density
function (PDF) takes values $\alpha\in(0,2]$, the skewness of the distribution
is modeled by the asymmetry parameter $\beta\in[-1,1]$. Indices $\alpha$ and
$\beta$ classify the type of stable distributions up to translations and
dilations \cite{janicki1994,janicki1996}. Two other parameters of scaling
$\sigma\in(0,\infty)$ and location $\mu\in(-\infty,\infty)$ can vary, although
replacing $\zeta-\mu$ and $\sigma \zeta$ in the original coordinates, shifts the
origin and rescales the abscissa without altering function
$L_{\alpha,\beta}(\zeta)$. For simplicity, we will restrict here to strictly
stable L\'evy noises with $\mu=0$ \cite{janicki1994,janicki1996}. Generally, for
$\beta=\mu=0$ PDFs are symmetric while for $\beta=\pm1$ and $\alpha\in(0,1)$
they are totally skewed, i.e., $\zeta$ is always positive or negative only,
depending on the sign of asymmetry parameter $\beta$ (cf.
Fig.~\ref{fig:density}). Asymptotically, for $\zeta\rightarrow\infty$ with
$\alpha<2$, stable PDFs behave as $p(\zeta)\propto |\zeta|^{-(\alpha+1)}$
causing divergence of moments $\langle\zeta^{\nu}\rangle=\infty$ for $\nu >
\alpha$. The asymmetry is reflected in a biased distribution
$\lim_{\zeta\rightarrow\infty}
 \frac{P(Z > \zeta)}{P(|Z| > \zeta)}=\frac{1+\beta}{2}$.

Equivalent to the stochastic ordinary differential Eq.~(\ref{eq:lang}) is a
fractional Fokker-Planck equation
\cite{fogedby1998,metzler1999,yanovsky2000,schertzer2001,dubkov2005} (FFPE) for the
probability distribution function
\begin{eqnarray}
\frac{\partial p(x,t)}{\partial t} & = & \frac{\partial}{\partial t} \int^{\infty}_{-\infty}\frac{dk}{2\pi}\phi(k,t)\e^{-ikx} \nonumber\\
 & = & \int^{\infty}_{-\infty}\frac{dk}{2\pi}\phi(k,t)
\e^{-ikx} \nonumber\\
& & \times
 \left[i\mu k-\sigma^{\alpha}|k|^{\alpha} +i\beta\sigma^{\alpha}k |k|^{\alpha-1} \tan \frac{\pi\alpha}{2} \right] \nonumber \\
& = &
- \sigma^{\alpha} (-\Delta)^{\alpha/2} p(x,t) \nonumber\\
& & -
 \sigma^\alpha\beta \tan\frac{\pi\alpha}{2}\frac{\partial}{\partial x}(-\Delta)^{(\alpha-1)/2} p(x,t)
\label{eq:sffpe}
\end{eqnarray}
with $p(x,t)$ representing the probability density functions for finding a
particle at time $t$ in the vicinity of $x$,
$\phi(k,t)=\left\langle\exp[ikx(t)]\right\rangle=\left\langle
\exp\left[ik\int^t_0 \zeta(s)ds\right]\right\rangle$ standing for the
characteristic function of the stable process and $-\Delta^{\alpha} f(x)$
denoting fractional Laplacian \cite{jespersen1999,metzler2000} $-
(\Delta)^{\alpha/2}f(x)=\mathcal{F}^{-1}\left(|k|^{\alpha}\hat{f}(k)\right)$,
with $\alpha=2$ corresponding to the standard Brownian diffusion case. The
addition of the potential force $-V'(x)$ to Eq.~(\ref{eq:lang}) adds the
classical drift term $\frac{\partial}{\partial x} \left[ V'(x)p(x,t) \right]$ to
Eq.~(\ref{eq:sffpe}). In the approach presented herein, instead of solving
Eq.~(\ref{eq:sffpe}), information on the system is drawn from the statistics of
numerically \cite{janicki1994,janicki1996,dybiecphd} generated trajectories
satisfying the generalized Langevin equation~(\ref{eq:lang}). At a single
trajectory level sampled from the stochastic dynamic study of the problem, the
initial condition for Eq.~(\ref{eq:lang}) has been set to
$x(0)\stackrel{\mathrm{d}}{=}\mathcal{U}[-1,1]$, i.e., initial position of the
particle is drawn from the uniform distribution over the interval $[-1,1]$. For
simplicity, Eq.~(\ref{eq:lang}) has been studied in dimensionless variables
\cite{chechkin2002,dybiec2007d} with additionally setting $\sigma=1$. The time
independent potential $V(x)$ is assumed to be of the form $V(x)=x^4/4$ which
guarantees the confinement of the trajectory $x(t)$ within the potential well
\cite{chechkin2002,chechkin2003,chechkin2004} leading to a finite variance of
the stationary PDF. For the general $\alpha$-stable driving and the quartic
potential the stationary PDF fulfills
\begin{eqnarray}
\frac{\partial^3 \hat{P}(k)}{\partial k^3}=\mathrm{sgn} k|k|^{\alpha-1}\hat{P}(k)-i \beta \tan\frac{\pi \alpha}{2}|k|^{\alpha-1}\hat{P}(k)
\label{eq:sffpes}
\end{eqnarray}
where $\hat{P}(k)$ stands for a Fourier transform of the stationary PDF, $P(x)=\lim_{t\rightarrow \infty} P(x,t)$.
Analytical solutions of
Eq.~(\ref{eq:sffpes}) can be readily obtained for a
Gaussian case $\alpha=2$: $P(x)\propto \exp(-x^4/4)$ and for a
Cauchy additive noise $\alpha=1$: $P(x)=1/[\pi(1-x^2+x^4)]$
\cite{chechkin2002,chechkin2003,chechkin2004}. They display a
perfect agreement with profiles of PDFs obtained by numerical simulations of
Eq.~(\ref{eq:lang})
performed with the Janicki--Weron algorithm
\cite{janicki1994,janicki1996,weron1995,weron1996}.

%%%%%%%%%%%%%%%%%%%%%%%%%%%%%%%%%%%%%%%%%%%%%%%%%%%%%%%%%%%%%%%%%%%%%%%%%%%%%%%%%%%%%%%%%
%%
%% RESULTS
%%
\section{Results}
Numerical results were constructed for a time step of integration $\Delta
t=10^{-3}$, simulation length $\Tm=10$ with an overall statistics of $N=5\times
10^4$ realizations. To check whether results are influenced by the length of
simulations, results for various $\Tm$ ($\Tm=10$, $\Tm=15$) were compared
showing consistency of estimated PDFs for both values. Further details on values
of parameters are included in the text underlying the figures. Numerical
examination of Eq.~(\ref{eq:lang}) allows for construction of PDF estimators for
the whole range of parameters $\alpha$ and $\beta$. Likewise, by direct
integration of  Eq.~(\ref{eq:lang}) it is also possible to investigate time-
dependent PDFs \cite{dybiecphd} and noise-induced bimodality of the probability
distribution \cite{chechkin2002,chechkin2003,chechkin2004,dybiecphd}.

%%%%%%%%%%%%%%%%%%%%%%%%%%%%
%%
%% gaus and cauchy
%%
\begin{figure}[!ht]
\includegraphics[angle=0, width=6.5cm]{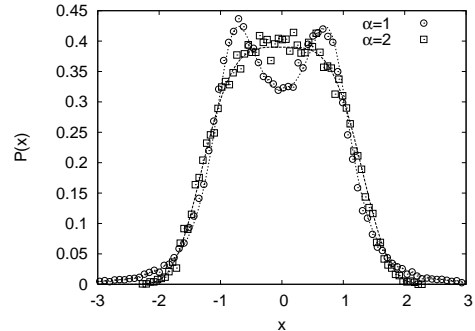}
\caption{Stationary probability density functions (PDFs) for $\alpha=2$
(Gaussian case, unimodal distribution) and $\alpha=1.0$ (Cauchy case, bimodal
distribution) with corresponding analytical solutions. Numerical results were
constructed for $\Delta t=10^{-3}$, $\Tm=10$ and averaged over $N=5\times10^4$ realizations.}
\label{fig:gausandcauchy}
\end{figure}

\begin{figure}
\begin{center}
\includegraphics[angle=0,width=8cm,height=5cm]{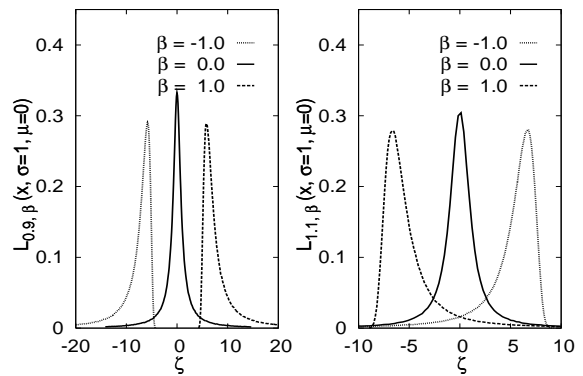}
\caption{Sample $\alpha$-stable PDFs with $\alpha = 0.9$ (left panel) and
$\alpha = 1.1$ (right panel). For $\beta = 0$ distributions are symmetric and
become asymmetric for $\beta=\pm 1$. The support of the densities for the fully
asymmetric cases with $\beta=\pm 1$ and $\alpha < 1$ (left panel) assumes only
negative values for $\beta = -1$ and only positive values for $\beta = 1$. Note
the differences in the positions of the maxima for $\alpha<1$ and $\alpha>1$.}
\label{fig:density}
\end{center}
\end{figure}

\begin{figure}[!ht]
\includegraphics[angle=0, width=8.0cm]{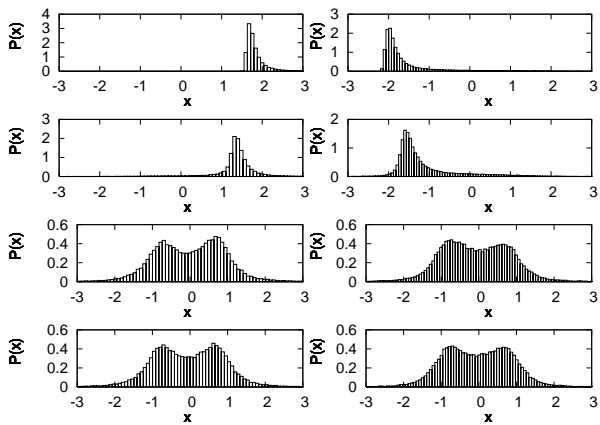}
\caption{Stationary probability distributions $p(x)$ for $\alpha=0.9$ (left
column) and $\alpha=1.1$ (right column). Various rows correspond to the various
values of $\beta$, starting from the top panel: $\beta=1$ (top row),
$\beta=0.5$, $\beta=0.01$, $\beta=0$ (bottom row). Due to symmetry, results for
negative $\beta$ can be obtained by the reflection of results for
$\beta\geqslant0$ along $x=0$ line. Results were constructed for $\Delta t=10^{-4}$, $\Tm=10$ and averaged over $N=5\times 10^4$ realizations.}
\label{fig:stationary}
\end{figure}

The change of $\alpha$ (for $\beta\neq0$) from values greater than 1 to values
smaller than 1 results in change of location of a modal value of stable
densities, see left vs. right panel of Fig.~\ref{fig:density}. The shift of
modal values is reflected in properties of stationary states. For $\alpha>1$
with $\beta<0$ the modal value is located for $x>0$ (right panel of
Fig.~\ref{fig:stationary}) while for $\alpha<1$ with $\beta<0$ it shifts to the
negative interval $x<0$ (left panel of Fig.~\ref{fig:stationary}). For $\beta<0$
modal values are located on the opposite side of the origin than for $\beta>0$.
Therefore, the position of the modal value can be moved from the positive to
negative real lines by change of $\alpha$ while $\beta$ is kept constant (left
vs. right panel of Fig.~\ref{fig:stationary}) or by change of $\beta$ to $-
\beta$ with a preset value of $\alpha$ (cf. different rows in
Fig.~\ref{fig:stationary}).

%%%%%%%%%%%%%%%%%%%%%%%%%%%%
%%
%% stable densities
%%

%%%%%%%%%%%%%%%%%%%%%%%%%%%%
%%
%% stationary m4
%%

\begin{figure}[!ht]
\includegraphics[angle=0, width=8.0cm,height=5cm]{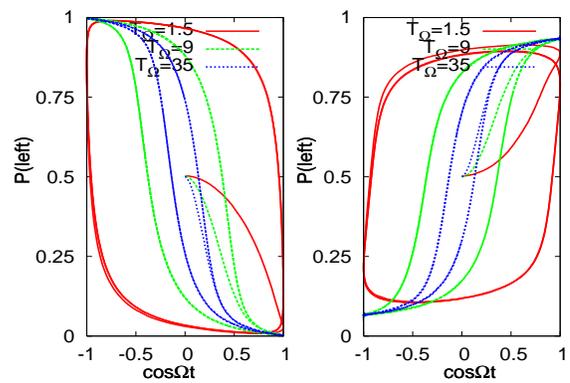}
\caption{(Color online) Hysteresis loops for $\alpha=0.9$ (left panel) and
$\alpha=1.1$ (right panel). Different lines correspond to the different values
of $T_\Omega$, amplitude of $\beta$ $\betam=1.0$. Results were constructed for
$\Delta t=10^{-3}$ and averaged over $5\times 10^4$ realizations. Various loops
corresponds to different values of the driving period $T_\Omega$: $T_\Omega=35$,
$T_\Omega=9$ and $T_\Omega=1.5$ (from the inside to outside).}
\label{fig:betahistereza}
\end{figure}

The fact that changes in $\beta$ can change the location of the modal value
suggests that periodic changes in this parameter can lead to a phenomenon
resembling dynamical hysteresis \cite{gudowska2004,juraszek2005,pustovoit2006}.
In order to register a hysteretic behavior of the system, we have performed an
analysis of trajectories of Eq.~(\ref{eq:lang}) based on a two-state
approximation. For that purpose, we have defined an occupation probability of
being in left/right state according to
\begin{eqnarray}
p_t(\mathrm{left}) & = & \mathrm{prob}\{x(t)<0\}=\int_{-\infty}^0p(x,t)dx \nonumber \\
& =& 1-p_t(\mathrm{right}).
\end{eqnarray}
Transition between the states is induced by time dependent asymmetry parameter
$\beta$ which is periodically modulated over time, i.e., $\beta=\betam\cos\Omega
t=\betam \cos \frac{2\pi}{T_\Omega}$. In the Fig.~\ref{fig:betahistereza} values
of $p(\mathrm{left})$ for various $T_\Omega$ (with $\betam=1$) are presented.
Stable random variables for $\alpha<1$ with $|\beta|=1$ take only
positive/negative values, depending on the sign of $\beta$. Therefore higher
level of saturation is observed for $\alpha=0.9$ (left panel) than for
$\alpha=1.1$ (right panel), i.e., when $|\beta|=1$ probability of being in the
left/right state for $\alpha=0.9$ is higher than for $\alpha=1.1$ (right panel).
The direction of the hysteresis loop is a direct consequence of the fact that
changes in $\beta$ move modal values from positive/negative real line to the
negative/positive real line. Due to the initial condition imposed on $x(0)$, the
starting point for each hysteresis loop is (0,0.5) and the first part of the
curve describes approaching to the proper hysteresis loop. With decreasing
$\Omega$ (increasing $T_\Omega$) area of hysteresis loop decreases because the
system has more time for relaxation and response to the changes of $\beta$. For
large $\Omega$ (small $T_\Omega$) the response of the system is more delayed and
consequently area of the hysteresis loop is larger.

%%%%%%%%%%%%%%%%%%%%%%%%%%%%
%%
%% betahistereza
%%

\begin{figure}[!ht]
\includegraphics[angle=0, width=8.0cm, height=5cm]{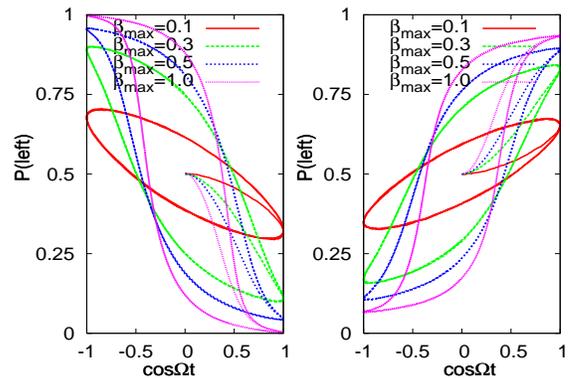}
\caption{(Color online) Hysteresis loops for $\alpha=0.9$ (left panel) and
$\alpha=1.1$ (right panel). Different lines correspond to the different values
of $\betam$, driving period $T_\Omega=9$. Results were constructed for $\Delta
t=10^{-3}$ and averaged over $5\times 10^4$ realizations. Various loops
corresponds to different values of $\betam$: $\betam=1.0$, $\betam=0.5$,
$\betam=0.3$ and $\betam=0.1$ (from the top to bottom).}
\label{fig:betamaxhistereza}
\end{figure}

Finally the influence of $\betam$ on the shape of hysteresis loop has been
examined. In Fig.~\ref{fig:betamaxhistereza} hysteresis loops for various
$\betam$ (with $T_\Omega=9$) are presented. Smaller $\betam$ makes stable
distribution less skewed and as a consequence, less probability mass becomes
located in the left state and the effect of saturation is less visible.
Furthermore, with decreasing $\betam$, loops become more oval and finally for
$\betam=0$ the hysteresis phenomenon disappears.

%%%%%%%%%%%%%%%%%%%%%%%%%%%%
%%
%% betamax histereza
%%

%%%%%%%%%%%%%%%%%%%%%%%%%%%%%%%%%%%%%%%%%%%%%%%%%%%%%%%%%%%%%%%%%%%%%%%%%%%%%%%%%%%%%%%%%
%%
%% SUMMARY & CONCLUSIONS
%%
\section{Summary and Conclusions}

The modulation of stable noise parameters modify shape of stationary densities
corresponding to Eq.~(\ref{eq:lang}). In particular skewed noise characterized
by nonzero asymmetry parameter $\beta$ can induce asymmetry of stationary states
in symmetric potentials. Furthermore, totally skewed stable noises ($\alpha<1$
with $|\beta|=1$) could move the probability mass to one side of the $x$-axis
making stationary states totally skewed. In such situations, the whole
probability mass can be located on the left hand side or on the right hand side
of the origin $x=0$ depending on the sign of the asymmetry parameter $\beta$.
For $\alpha>1$ with a nonzero $\beta$, stable noises still are asymmetric.
Nevertheless, even $|\beta|=1$ is not sufficient to induce totally skewed
stationary states. Consequently, dynamical hysteresis loops induced by cyclic
variation of $\beta$ for $\alpha<1$ and $\alpha>1$ are characterized by various
level of saturation, see left vs. right panel of Fig.~\ref{fig:betahistereza}.

The stationary densities depicted in Fig.~\ref{fig:stationary} are recorded in
the system described by Eq.~(\ref{eq:lang}) which is subjected to the action of
stable noises. The very same stationary densities can be also observed in the
equilibrium system perturbed by the white Gaussian noise, i.e., $\dot{x}(t)=-
\Veff'(x,t)+\xi(t)$, where $\Veff(x,t)=-\log[p(x,t)]$ and $\langle \xi(t)\xi(t')
\rangle=\delta(t-t')$. The dynamical hysteresis detected in the model described
by Eq.~(\ref{eq:lang}) emerges as a consequence of periodical modulation of the
asymmetry parameter $\beta$. In the effective potential model, periodical
modulation of the asymmetry parameter corresponds to the periodical modulation
\cite{jung1990} of the effective potential $\Veff(x)$. Nevertheless both models
are not fully equivalent. They have the same one-point densities $p(x,t)$, while
other characteristic of the process $\{x(t)\}$ are distinct \cite{dybiec2007d}.
Therefore, examination of stationary densities itself is not a fully conclusive
method for discrimination of underlying types of noises.

The dynamical hysteresis can be also observed in the generic double well
potential model subject to the joint action of the deterministic periodic
modulation and stochastic $\alpha$-stable fluctuations \cite{dybiecphd}. In such
a case, however, the shape of the dynamical hysteresis loop is affected both by
the character of the noise and by the type of periodic modulation
\cite{dybiecphd}.

%%%%%%%%%%%%%%%%%%%%%%%%%%%%%%%%%%%%%%%%%%%%%%%%%%%%%%%%%%%%%%%%%%%%%%%%%%%%%%%%%%%%%%%%%
%%
%% ACKNOWLEDGMENTS
%%
\begin{acknowledgments}
The research has been supported by the Marie Curie TOK COCOS grant (6th EU
Framework Program under Contract No. MTKD-CT-2004-517186). Computer simulations
have been performed at the Academic Computer Center CYFRONET AGH, Krak\'ow.
Additionally, BD acknowledges the support from the Foundation for Polish Science
and the hospitality of the Humboldt University of Berlin and the Niels Bohr
Institute (Copenhagen).
\end{acknowledgments}

%%%%%%%%%%%%%%%%%%%%%%%%%%%%%%%%%%%%%%%%%%%%%%%%%%%%%%%%%%%%%%%%%%%%%%%%%%%%%%%%%%%%%%%%%
%%
%% BIBLIOGRAPHY
%%
%\bibliography{bibliography}

\end{document}